\documentclass[10pt,twocolumn]{article}
\usepackage{amsmath}
\usepackage{amsfonts}
\usepackage{amssymb}
\usepackage{graphicx}

\begin{document}

\title{Van der Waal's gas equation for an adiabatic process and its Carnot engine efficiency}
\author{Kiran S. Kumar, Aravind P. Babu and M. Ponmurugan* \\
Department of Physics, School of Basic and Applied Sciences, \\
Central University of Tamilnadu, Tiruvarur 610 005, Tamilnadu, India.} 
\date{}
\maketitle

\begin{abstract}
There has been many studies on gases which obeys Van der Waal's 
equation of state. However there is no specific and direct studies of 
Van der Waal's gas which undergoes adiabatic processes are available
in the undergraduate text books and also in literature. In an adiabatic process 
there is no heat energy exchange between the system and its surroundings.
In this article, we  find that the Van der Waal's equation for the adiabatic process
as $\left(P+\frac{n^2a}{V^2}\right) \left(V-nb\right)^{\Gamma}=\mbox{constant}$,
where $P$ is the pressure, $V$ is the volume, $n$ is the number of moles 
of the Van der Waal's gas, $a$ and $b$ are Van der Waal's constant and 
$\Gamma$ is a factor which relates the specific heat at constant pressure and at constant volume. 
We use this relation explicitly and obtained the efficiency of a  
Carnot engine whose working substance obeys Van der Waal's equation of state. 
Our simplest approach  may provide clear idea to the undergraduate students that 
$\Gamma$ is different from $\gamma$ of the ideal gas  for an adiabatic process. We also 
shown that the efficiency of the Carnot engine is independent of the working substance.    
\end{abstract}


\section{Introduction}
In thermodynamics, heat and work are the form of energy transfer across the boundary between the system and its surroundings. If the
boundary forbids the flow of heat energy between the system and the surroundings then the thermodynamic process which changes the 
state of the system is called as adiabatic process. If the energy transfer takes at a fixed
system temperature then such a process is called as isothermal process. 
All heat engines make use of the mechanism of converting heat $(Q)$ in to work $(W)$, without involving any resultant change in the state of the system. It is a series of processes taking place in a cyclic manner in which the engine or system will be returned to its initial state. 

During each of the processes, there may be a heat energy flow between the system and its surroundings. This comprises of a hot reservoir and a cold reservoir, which will be maintained at  constant temperatures, and the working substance which exchanges heat. Therefore, the engine is said to operate between these two reservoirs.
During a part of cycle performed by the working substance in an engine, some heat $Q_H$ is absorbed 
from the hotter reservoir and a smaller amount of heat $Q_L$ is rejected to the cooler reservoir during the another part of cycle.  
In a cyclic process, the system returns to the initial state with no change in internal energy ($\Delta U=0$). 
So, from the first law of thermodynamics $Q_{net}-W_{net}=0$, where $Q_{net}=Q_H-Q_L$ is the net heat exchanged between the
system and the surroundings and $W_{net}=Q_{net}= Q_H-Q_L$ is the total or net work performed on the system. 
The engine efficiency $\eta$ is defined as \cite{zeman,sear,sch},
\begin{eqnarray}
\eta=\frac{\mbox{Net Work}}{\mbox{Heat  absorbed}}=\frac{Q_H-Q_L}{Q_H} =1-\frac{Q_L}{Q_H}.
\label{eff}
\end{eqnarray}

\subsection{Carnot Cycle}
Carnot proposed a thermodynamic cycle called as Carnot cycle which is a set of equilibrium
reversible processes any thermodynamic system can perform \cite{zeman,sear}. Initially the system or working 
substance in Carnot cycle 
is imagined to be in thermal equilibrium with a reservoir at lower temperature $T_{L}$. 
Four processes are then performed in the following order:
\begin{enumerate}
\item A reversible adiabatic process is performed in such a direction that the temperature rises to that of hotter 
reservoir temperature $T_{H}$.
However, no energy in the form of heat flows in to or out of the system.
\item The working substance is maintained in contact with the reservoir at $T_{H}$ and a reversible isothermal process is performed in such a direction to the extent that $Q_{H}$ is absorbed from the reservoir. In this case the system is kept at a fixed 
reservoir temperature.
\item A reversible adiabatic process is performed in a direction opposite to that of the first process with no heat exchange.
This is done until the temperature reaches $T_{L}$, temperature of the cooler reservoir.
\item  A reversible isothermal process opposite to the direction of the second process is performed until the working substance reaches the initial state and $Q_{L}$ is rejected by the working substance. In this case the system is kept at a fixed 
reservoir temperature $T_L$.
\end{enumerate}
Thus an engine in Carnot cycle operates between two reservoirs in a particular simple way. 
All the absorbed heat enters the system at a constant high temperature, 
namely that of a hotter reservoir. Also, all the rejected heat leaves the system 
at a constant low temperature, that of a cooler reservoir. Since all processes are 
reversible, the Carnot engine is a reversible engine. 
An engine which operates in this cycle is called as Carnot engine
whose efficiency is always greater than any engines operated by the cycle 
other than Carnot cycle. 
Any hypothetical engine operated in this cycle 
is said to be ideal if it gives $100\%$ efficiency.    
Since it is fact of experience that some heat is always rejected to 
cooler reservoir, the efficiency of actual engine is  always less 
than the ideal and the Carnot engine \cite{zeman,sear,sch}.

Considering a working substance in a Carnot engine as an ideal gas, with no intermolecular
interactions, the gas satisfies the ideal gas equation of state $PV=nRT$. Where $P$
is the pressure, $V$ is the volume, $T$ is the temperature of the system,
$n$ is the number of moles and $R$ is the universal gas constant. 
The thermal efficiency of a Carnot engine whose working substance 
is an ideal gas is given by \cite{sear,sch}
\begin{eqnarray}
\eta &=&1-\frac{T_L}{T_H}.
\label{effTi}
\end{eqnarray} 
Thus, a Carnot engine absorbing $Q_H$ amount of heat from the reservoir at  higher emperature $T_H$ 
and rejecting $Q_L$ amount of heat to the reservoir at  lower temperature $T_L$ has an efficiency 
that is independent of the nature of the working substance.
The above result is obtained by using the ideal gas equation of 
state for the isothermal process $PV=\mbox{constant}$ and the ideal gas equation for the 
adiabatic process $PV^{\gamma}= \mbox{constant}$. Here, $\gamma$ is the 
ratio of the specific heat capacity at constant pressure to the  
specific heat capacity at constant volume. Further studies using  
various gas equations \cite{old1,cvan,arbit1,arbit2} 
also showed that the efficiency of 
a Carnot cycle is independent of working substance and 
depends only on temperatures of reservoirs.

\subsection{Van der Waal's gas}

By considering the intermolecular interactions, Van der Waal proposed the 
equation of state for a real gas which is given by\cite{sear,sch,cvan}
\begin{eqnarray}
\left( P+\frac{n^{2}a}{V^2} \right)(V-nb)=nRT,
\label{van}
\end{eqnarray}
where $a$ and $b$ are the Van der Waal's constants.  
There has been several studies for Van der Waal's 
equation of state \cite{cvan,arbit1}. However,
there is no specific and direct studies available in 
undergraduate text books and also in literature for the equation of 
Van der Waal's gas subjected to an adiabatic process.
In this paper we explicitly find that  the Van der Waal's equation 
for an adiabatic process as
\begin{eqnarray}
\left( P+\frac{n^{2}a}{V^2} \right)\left (V-nb \right )^{\Gamma}=\mbox{constant},
\label{vanadi}
\end{eqnarray}
where $\Gamma$ is a factor which relates the specific heat at constant pressure and at constant volume. 
We use the above relation and obtained the relation between the specific
heat capacity at constant pressure and at constant volume.
As similar to ideal gase approach we also used the above relation directly 
and showed that the efficiency of a Carnot cycle is independent 
of the working substance and depends only on the temperature
of the reservoirs.

\section{Van der Waal's gas equation for an Adiabatic process}

The entropy of pure substance can be considered as a function of any two 
variable $T$ and $V$ as 
\begin{eqnarray}
S&=& S(T,V) \\ \nonumber
dS &=& \Big(\frac {\partial S }{\partial T} \Big)_V dT + 
\Big(\frac {\partial S }{\partial V} \Big)_T dV.
\label{ds1}
\end{eqnarray}
Multiply both sides by $T$ we get,
\begin{eqnarray}
TdS&=& T\Big(\frac {\partial S }{\partial T} \Big)_V dT + 
T\Big(\frac {\partial S }{\partial V} \Big)_T dV.
\label{ds2}
\end{eqnarray}
Since $dQ=nC_V dT$ and $TdS=dQ$ for a reversible isochoric process, from the 
above equation one can obtain 
\begin{eqnarray}
T\Big(\frac {\partial S }{\partial T} \Big)_V &=& nC_V, 
\label{ds3}
\end{eqnarray}
where $C_V$ is the specific heat capacity at constant volume.
According to Maxwell's third thermodynamic relation
\begin{eqnarray}
\Big(\frac {\partial S }{\partial V} \Big)_T &=& \Big(\frac {\partial P }{\partial T} \Big)_V.
\label{mrel}
\end{eqnarray}
Therefore Eq.(\ref{ds2}) becomes 
\begin{eqnarray}
TdS&=& nC_V dT + 
T\Big(\frac {\partial P }{\partial T} \Big)_V dV.
\label{ds4}
\end{eqnarray}
For an adiabatic process $dQ=TdS=0$, hence
\begin{eqnarray}
nC_V dT &=&-T\Big(\frac {\partial P }{\partial T} \Big)_V dV.
\label{ds5}
\end{eqnarray}
Using Eq.(\ref{van}), one can obtain
\begin{eqnarray}
\Big(\frac {\partial P }{\partial T} \Big)_V &=& \frac{nR}{\big( V-nb \big)}.
\label{ds6}
\end{eqnarray}
Therefore Eq.(\ref{ds5}) becomes
\begin{eqnarray}
nC_V dT &=&-T \frac{nR}{\big( V-nb \big)} dV  \\ \nonumber
\frac{1}{T} \ dT &=&- \frac{R}{C_V} \frac{dV}{\big( V-nb \big)}. 
\label{ds7}
\end{eqnarray}
Integrating the above equation we get
\begin{eqnarray}
\ell  n \ T &=& - \frac{R}{C_V} \  \ell n \ \big( V-nb \big) +  \ell n \ z,
\label{lnT1}
\end{eqnarray}
where $\ell n z$ is an integrating constant. Rearranging the above equation
one can get
\begin{eqnarray}
 T \big( V-nb \big)^{\frac{R}{C_V}} &=& z.
\label{lnT2}
\end{eqnarray}
Combining Eq.(\ref{van}) and Eq.(\ref{lnT2}) one can obtain 
Van der Waal's gas equation for an adiabatic process as
\begin{eqnarray}
\left( P+\frac{n^{2}a}{V^2} \right)(V-nb)^{\Gamma}&=& K,
\label{vanadi1}
\end{eqnarray}
where $\Gamma=\frac{R}{C_V}+1$ and $K=nRz=$ a constant. It should be noted
that, for real gas $\Gamma \neq  \gamma=\frac{C_P}{C_V}$ of an ideal gas,
where $C_P$ is the specific heat capacity at constant pressure. Thus,
it would be interesting to obtain the relation between the $C_P$ and $C_V$
of the Van der Waal's gas as follows.

\subsection{Relation between $C_P$ and $C_V$ for Van der Waal's gas}

The internal energy of a given system can be considered as a function of any two 
variable $T$ and $V$ as
\begin{eqnarray}
U &=& U(T,V) \\ \nonumber
dU &=& \Big(\frac {\partial U }{\partial T} \Big)_V dT + 
\Big(\frac {\partial U }{\partial V} \Big)_T dV.
\label{du1}
\end{eqnarray}
Using the above equation, the first law of thermodynamics in an 
infinitesimal form $dQ=dU+dW$ with $dW=-PdV$, can be written as
\begin{eqnarray}
dQ &=& \Big(\frac {\partial U }{\partial T} \Big)_V dT + 
\Big(\frac {\partial U }{\partial V} \Big)_T dV - PdV.
\label{dq1}
\end{eqnarray}
In our  study we have used the sign convention that the 
work done on the system is taken as positive and the
work done by the system is taken as negative.  
For constant volume $dV=0$, the above equation becomes,
\begin{eqnarray}
nC_V &=& \Big(\frac {\partial U }{\partial T} \Big)_V.
\label{dq2}
\end{eqnarray}
Since $dQ/dT=nC_V$ for constant volume and $dQ/dT=nC_P$ for constant 
pressure, Eq.(\ref{dq1}) can be rewritten as
\begin{eqnarray}
dQ =nC_V dT + \Big(\frac {\partial U }{\partial V} \Big)_T dV - PdV \\
nC_P dT = nC_V dT + \Big(\frac {\partial U }{\partial V} \Big)_T dV - PdV \\
n(C_P-C_V)= \Big\{\Big(\frac {\partial U }{\partial V} \Big)_T - P \Big\} \frac{dV}{dT}.
\label{nc1}
\end{eqnarray}
From the first thermodynamic potential $dU=TdS+PdV$ and using  Eq.(\ref{du1}) and Eq.(\ref{mrel}) at constant temperature one can obtain
\begin{eqnarray}
\Big(\frac {\partial U }{\partial V} \Big)_T &=&
 T \Big(\frac {\partial S }{\partial V} \Big)_T + P \\ 
   &=& T \Big(\frac {\partial P }{\partial T} \Big)_V + P . 
\label{rel1}
\end{eqnarray}
Then Eq.(\ref{nc1}) becomes
\begin{eqnarray}
n(C_P-C_V) &=& T \Big(\frac {\partial P }{\partial T} \Big)_V \frac{dV}{dT}.
\label{nc3}
\end{eqnarray}
If $V$ is a function of $T$ and $P$, then change in $V$ for 
a constant pressure is 
\begin{eqnarray}
dV=\Big(\frac {\partial V }{\partial T} \Big)_P dT +
\Big(\frac {\partial V }{\partial P} \Big)_T dP.
\label{vol1}
\end{eqnarray}
\begin{eqnarray}
\frac{dV}{dT}=\Big(\frac {\partial V }{\partial T} \Big)_P. 
\label{vol2}
\end{eqnarray}

From Eq.(\ref{van}) on can obtain 
\begin{eqnarray}
\Big(\frac {\partial P }{\partial T} \Big)_V=\frac{nR}{\big( V-nb \big)},
\label{rs1}
\end{eqnarray}
\begin{eqnarray}
\Big(\frac {\partial V }{\partial T} \Big)_P=
nRV^3 \big( V-nb \big) {\it G}^{-1}, 
\label{rs2}
\end{eqnarray}
where $G=V^3nRT-2n^2a \big( V-nb \big)^2$. \\
Therefore Eq.(\ref{nc3}) can be written as
\begin{eqnarray}
C_P-C_V = \frac{R}{f_v},
\label{cfin1}
\end{eqnarray}
where
\begin{eqnarray}
f_v = \Big\{1- \frac{2na} {V^3RT} \big( V-nb \big)^2 \Big\}.   
\label{cfin2}
\end{eqnarray}
We have obtained the relation between the $C_P$ and 
$C_V$ of Van der Waals gas.
Divide Eq.(\ref{cfin1}) throughout by $C_V$ we get
\begin{eqnarray}
\frac{C_P}{C_V}=1+ \frac{R}{f_v C_V}=1+ \frac{\Gamma-1}{f_v}.
\label{may1}
\end{eqnarray}
Thus we have finally  obtained the relation between $\gamma$ of ideal 
gas and $\Gamma$ of Van der Waal's gas as
\begin{eqnarray}
\gamma=1+ \frac{\Gamma-1}{f_v}.
\label{may2}
\end{eqnarray}

There has been few studies to find out the efficiency of the Carnot cycle 
whose working substance is different from ideal gases \cite{cvan,arbit1,arbit2}. 
In what follows, we employ simple approach to find out the efficiency of the Carnot engine 
whose working substance obeys the Van der Waal's equation of state.

\section{Carnot engine efficiency for Van der Waal's gas}

Efficiency of the  Carnot engine, $\eta$, is defined as the ratio of the net work done to 
the heat absorbed in the Carnot cycle. As discussed earlier, this reversible cycle 
consist of four processes such as
i) adiabatic compression
ii) isothermal expansion
iii) adiabatic expansion  and 
iv) isothermal compression.
In order to find out $\eta$, we calculate the total work done 
during the Carnot cycle as follows.


\subsection{Work done in an adiabatic compression}
In this process, the volume changes from $V_1$ to $V_2$ $(V_1 > V_2)$, 
the pressure changes from $P_1$ to $P_2$ and the temperature from $T_L$ to $T_H$. 
There is no heat exchange between the system and the surroundings.  
The work done during this process $W_1$ is given by,
\begin{eqnarray}
W_1=- \int_{V_1}^{V_2}PdV.
\label{wdef}
\end{eqnarray}
The Van der Waal's equation for an adiabatic process obtained in Eq.(\ref{vanadi1}) as
\begin{eqnarray}
\left(P+\frac{n^2a}{V^2}\right)\left(V-nb\right)^{\Gamma}= K.
\label{adivank}
\end{eqnarray}
\begin{eqnarray}
P=\frac{K}{(V-nb)^{\Gamma}}-\frac{n^2a}{V^2}.
\label{adivanp}
\end{eqnarray}
Substitute Eq.(\ref{adivanp}) in Eq.(\ref{wdef}) and integrating, we get
\begin{eqnarray*}
W_1= - \frac{K(V-nb)^{1-\Gamma}}{1-\Gamma}\Bigg\vert_{V_1}^{V_2} - \frac{n^2 a}{V}\Bigg\vert_{V_1}^{V_2},
\end{eqnarray*}
\begin{eqnarray*}
W_1= \frac{K(V_1 -nb)^{1-\Gamma}-K(V_2 -nb)^{1-\Gamma}}{1-\Gamma}-\frac{n^2 a}{V_2}+\frac{n^2 a}{V_1}.
\end{eqnarray*}
The Van der Waal's equation of state for the system in the initial state ($P_1, V_1$) at $T_{L}$ is
\begin{eqnarray}
\left(P_1+\frac{n^{2}a}{V_{1}^2}\right)(V_1-nb)=nRT_L
\label{adp1}
\end{eqnarray}
and for the system in the final state ($P_2, V_2$) at $T_{H}$ as
\begin{eqnarray}
\left(P_2+\frac{n^{2}a}{V_{2}^2}\right)(V_2-nb)=nRT_H.
\label{adp2}
\end{eqnarray}
From the Van der Waal's equation (Eq.\ref{adivank}) for the adiabatic process, 
we can relate the system in the initial state ($P_1, V_1,T_L$) and 
final state ($P_2, V_{2}, T_H$) as
\begin{eqnarray}
\left( P_1+\frac{n^{2}a}{V_1^2} \right)\left (V_1-nb \right)^{\Gamma}=\left( P_2+\frac{n^{2}a}{V_2^2} \right)\left (V_2-nb \right)^{\Gamma}. 
\label{adieq1}
\end{eqnarray}
Substitute Eq.(\ref{adp1}) and Eq.(\ref{adp2}) in the above equation, we get
\begin{eqnarray}
nRT_L (V_1-nb)^{\Gamma-1}=nRT_H (V_2-nb)^{\Gamma-1} \equiv K.
\end{eqnarray}
Therefore,
\begin{eqnarray}
K(V_1-nb)^{1-\Gamma}=nRT_L \\
K(V_2-nb)^{1-\Gamma}=nRT_H 
\end{eqnarray}
and hence the work done in an adiabatic compression becomes
\begin{equation}
W_1=\frac{nR(T_L -T_H)}{1-\Gamma}-\frac{n^2 a}{V_2}+\frac{n^2 a}{V_1}.
\end{equation}

\subsection{Work done in an isothermal expansion}
In this process, the system  undergoes volume expansion $V_{2}$  to $V_{3}$ and 
the pressure change from $P_2$ to $P_3 $, while the temperature remains 
constant at $T_H$. During this process the system absorbs $Q_H$ amount of heat energy 
from the hot reservoir, then the work done,
\begin{eqnarray}
W_2 = - \int_{V_{2}}^{V{3}} PdV.
\label{wdefise}
\end{eqnarray}
For a reservoir temperature $T_H$, Eq.(\ref{van}) can be written as,
\begin{eqnarray}
P=\frac{nRT_H}{V-nb}-\frac{n^2a}{V^2}
\end{eqnarray}
Substitute the above equation in Eq.(\ref{wdefise}) and integrating, 
we get 
\begin{eqnarray}
W_2=nRT_H \ell n \left(\frac{V_2-nb}{V_3-nb}\right)-\frac{n^2a}{V_3}+\frac{n^2a}{V_2}.
\end{eqnarray}
From the Van der Waal's equation of state (Eq.\ref{van}), 
we can relate the system in the initial state ($P_2, V_2$) and 
final state ($P_3, V_{3}$) at a fixed temperature $T_H$ as
\begin{eqnarray}
\left( P_2+\frac{n^{2}a}{V_2^2} \right)(V_2-nb)=\left( P_3+\frac{n^{2}a}{V_3^2} \right)(V_3-nb).
\label{isoeq1}
\end{eqnarray}
Substitute Eq.(\ref{isoeq1}) in Eq.(\ref{adieq1}), then
\begin{eqnarray}
\frac{P_1+\frac{n^{2}a}{V_1^2}}{P_3+\frac{n^{2}a}{V_3^2}}=\frac{(V_3-nb)(V_1-nb)^{-\Gamma}}{(V_2-nb)^{1-\Gamma}}.
\label{adiiso1}
\end{eqnarray}

\subsection{Work done in an adiabatic expansion}

As similar to adiabatic compression, the  heat exchange  in an adiabatic expansion is zero.
As the system expands from $V_3$ to $V_4$, the pressure changes from $ P_3$ to $P_4$, and the temperature 
changes from $T_H$ to $T_L$, then the work done $W_3$ during this process is given by,
\begin{equation}
W_3=-\int_{V_3}^{V_4}PdV.
\end{equation}
Substitute Eq.(\ref{adivanp}) for an adiabatic process in the above equation, we get
\begin{eqnarray*}
W_3=\frac{K(V_3 -nb)^{1-\Gamma}-K(V_4 -nb)^{1-\Gamma}}{1-\Gamma}-\frac{n^2 a}{V_4}+\frac{n^2 a}{V_3}.
\end{eqnarray*}
The Van der Waal's equation of state for the system in the initial state ($P_3, V_3$) at $T_{H}$ is
\begin{eqnarray}
\left(P_3+\frac{n^{2}a}{V_{3}^2}\right)(V_3-nb)=nRT_H
\label{adpe1}
\end{eqnarray}
and for the system in the final state ($P_4, V_4$) at $T_{L}$ as
\begin{eqnarray}
\left(P_4+\frac{n^{2}a}{V_{4}^2}\right)(V_4-nb)=nRT_L.
\label{adpe2}
\end{eqnarray}
From the Van der Waal's equation (Eq.\ref{adivank}) for the adiabatic process, 
we can relate the system in the initial state ($P_3, V_3,T_H$) and 
final state ($P_4, V_{4}, T_L$) as
\begin{eqnarray}
\left( P_3+\frac{n^{2}a}{V_3^2} \right)\left (V_3-nb \right)^{\Gamma}=\left( P_4+\frac{n^{2}a}{V_4^2} \right)\left (V_4-nb \right)^{\Gamma}.
\label{adieq2}
\end{eqnarray}
Substitute Eq.(\ref{adpe1}) and Eq.(\ref{adpe2}) in the above equation, we get
\begin{eqnarray}
nRT_H (V_3-nb)^{\Gamma-1}=nRT_L (V_4-nb)^{\Gamma-1} \equiv K.
\end{eqnarray}
Therefore,
\begin{eqnarray}
K(V_3-nb)^{1-\Gamma}=nRT_H \\
K(V_4-nb)^{1-\Gamma}=nRT_L 
\end{eqnarray}
and hence the work done in an adiabatic expansion becomes
\begin{equation}
W_3=\frac{nR(T_H -T_L)}{1-\Gamma}-\frac{n^2 a}{V_4}+\frac{n^2 a}{V_3}.
\end{equation}

\subsection{Work done in an isothermal compression}

In this process, the pressure changes from $P_4$ to $P_1$, the volume changes 
from $V_4$ to $V_1$ and $Q_L$ amount of heat energy rejected to a 
cold reservoir at constant temperature $T_L$.
Work done during this process $W_4$ is given by
\begin{eqnarray}
W_4 = -\int_{V_{4}}^{V{1}} PdV.
\label{wdefisc}
\end{eqnarray}
For a reservoir temperature $T_L$, Eq.(\ref{van}) can be written as,
\begin{equation}
P=\frac{nRT_L}{V-nb}-\frac{n^2a}{V^2}
\end{equation}
Substitute the above in equation in Eq.(\ref{wdefisc}) and integrating, 
we get 
\begin{equation}
W_4=nRT_L  \ell n \left(\frac{V_4-nb}{V_1-nb}\right)-\frac{n^2a}{V_1}+\frac{n^2a}{V_4}.
\end{equation}
From the Van der Waal's equation of state (Eq.\ref{van}), 
we can relate the system in the initial state ($P_4, V_4$) and 
final state ($P_1, V_{1}$) at a fixed temperature $T_L$ as
\begin{eqnarray}
\left( P_4+\frac{n^{2}a}{V_4^2} \right)(V_4-nb)=\left( P_1+\frac{n^{2}a}{V_1^2} \right)(V_1-nb).
\label{isoeq2}
\end{eqnarray}
Substitute Eq.(\ref{isoeq2}) in Eq.(\ref{adieq2}), then
\begin{eqnarray}
\frac{P_3+\frac{n^{2}a}{V_3^2}}{P_1+\frac{n^{2}a}{V_1^2}} =\frac{(V_1-nb)(V_3-nb)^{-\Gamma}}{(V_4-nb)^{1-\Gamma}}.
\label{adiiso2}
\end{eqnarray}

\subsection{Efficiency of the engine}
The net  work done $W_{net}$ for one complete cycle is 
$W_{net}=W_1+W_2+W_3+W_4$.
Therefore, the total work done for the Carnot cycle is given by 
\begin{eqnarray}
W_{net}&=&w_H+w_L,
\label{netw}
\end{eqnarray}
where
\begin{eqnarray}
w_H &=&nRT_H \ell n \left( \frac{V_2-nb}{V_3-nb} \right)  \\
w_L &=& nRT_L \ell n \left( \frac{V_4-nb}{V_1-nb} \right).
\label{wHL}
\end{eqnarray}

According to the first law of thermodynamics $Q_{net}-W_{net}=\Delta U$, where $Q_{net}=Q_H-Q_L$ is
the net heat energy exchange between the system and the reservoir and 
$\Delta U$ is the change in the internal energy.
Here we have used the sign convention that the heat energy flow in to
the system is taken as positive and the heat energy flow out of 
the system is taken as negative. For a cyclic process, the change in 
internal energy $\Delta U$ is zero. Therefore, 
the net heat energy exchange between the system and the reservoir in a given cycle is 
completely converted in to net work done  which is given by
\begin{eqnarray}
W_{net}&=&Q_H-Q_L.
\label{netq}
\end{eqnarray}
In the Carnot cycle, all the absorbed heat $Q_H$ enters the system 
at a constant high temperature $T_H$ and all the rejected heat $Q_L$ leaves
the system at a constant low temperature $T_L$. 
Hence, comparing Eq.(\ref{netw}) and Eq.(\ref{netq}), one can identify\footnote {It should be 
noted that $Q$ can also be obtained from other approaches. Using Eq.(\ref{rel1}) and Eq.(\ref{rs1}) 
for an isothermal process, Eq.(\ref{dq1}) becomes
\[dQ = T \Big(\frac {\partial P }{\partial T} \Big)_V  dV =\frac{nRT}{\big( V-nb \big)} dV. \]
Integrating the above equation from the initial volume $V_i$ to the final volume $V_f$, one can obtain the 
amount of heat transferred between the system and the sorrounding as 
\[Q = nRT \ell n \big(V-nb \big) \Bigg|_{V_i}^{V_f}.\] }
 
\begin{eqnarray}
Q_H&=& \Big| w_H \Big|=\Bigg| \ nRT_H \ell n \left(\frac{V_2-nb}{V_3-nb} \right)\ \Bigg| \\
-Q_L&=&  \Big| w_L \Big| = \Bigg| \ nRT_L \ell n \left(\frac{V_4-nb}{V_1-nb} \right) \ \Bigg|.
\end{eqnarray}
The efficiency of the engine obtained from Eq.(\ref{eff}) as 
\begin{eqnarray*}
\eta=1-\frac{Q_L}{Q_H}=1+\frac{T_L \ \ell n\left(\frac{V_4-nb}{V_1-nb}\right)}{T_H \  \ell n\left(\frac{V_2-nb}{V_3-nb}\right)}.
\end{eqnarray*}
The above equation can be rewritten as
\begin{eqnarray}
\eta&=&1-\frac{T_L \ \ell n\left(\frac{V_4-nb}{V_1-nb}\right)}{T_H \ \ell n\left(\frac{V_3-nb}{V_2-nb}\right)}. 
\label{etalast}
\end{eqnarray}
In order to simply the above equation, substitute Eq.(\ref{adiiso2}) in Eq.(\ref{adiiso1}) 
and rearranging, we get
\begin{equation*}
(V_1 - nb)^{1-\Gamma}(V_3 -nb)^{1-\Gamma} =(V_2 - nb)^{1-\Gamma}(V_4-nb)^{1-\Gamma}
\end{equation*}
\begin{equation*}
\Bigg(\frac{V_3 -nb}{V_2 -nb}\Bigg)^{1-\Gamma}= \Bigg(\frac{V_4 -nb}{V_1 -nb}\Bigg)^{1-\Gamma}.
\end{equation*}
So,
\begin{equation}
\Bigg(\frac{V_3 -nb}{V_2 -nb}\Bigg)= \Bigg(\frac{V_4 -nb}{V_1 -nb}\Bigg)
\end{equation}
Therefore, Eq.(\ref{etalast}) reduces to 
\begin{equation}
\eta = 1 - \frac{T_L}{T_H}.
\end{equation}
Thus, we used Van der Waal's gas as a working substance and obtained 
the efficiency of the Carnot engine which is independent 
of the working substance. 

\section{Conclusion}
In summary, we have obtained the Van der Waal's equation for the adiabatic process
as $\left(P+\frac{n^2a}{V^2}\right) \left(V-nb\right)^{\Gamma}=\mbox{constant}$. 
Our result explicitly shows that $\Gamma$ of the Van der Waal's gas 
is different from $\gamma$ of ideal gas for adiabatic process.   
This equation has been used directly in Carnot cycle 
for the adiabatic process and shown that the efficiency of the Carnot engine is 
independent of the working substance. For this calculation we have used the simple 
approach as similar to the ideal gas usually found in the undergraduate text books.
With this we  find out the efficiency of the Carnot engine whose working substance 
obeys Van der Waal's equation of state 
$\left(P+\frac{n^2a}{V^2}\right) \left(V-nb\right)=nRT$.
Using the above two equations, we have also shown that the alternative 
Van der Waal's equation for an adiabatic process is
$T(V-nb)^{\Gamma-1}= \mbox{constant}$. We can also write 
the above relation as
$\left(P+\frac{n^2a}{V^2}\right) T^{\frac{\Gamma}{1-\Gamma}}=\mbox{constant}$.

\paragraph*{}

*Corresponding author: ponphy@cutn.ac.in

\end{document}